\documentclass[]{agujournal2019}
\usepackage{url} 
\usepackage[finalnew]{trackchanges}
\usepackage{soul}

\draftfalse

\journalname{JGR: Space Physics}

\begin{document}

\newcommand{\apj} {The Astrophysical Journal}
\newcommand{\apjl} {The Astrophysical Journal Letter}
\newcommand{\jgr} {Journal of Geophysical Research Letter}
\newcommand{\grl} {Geophysical Research Letter}
\newcommand{\solphys} {Solar Physics}
\newcommand{\ssr} {Space Science Reviews}
\newcommand{\aap }{Astronomy \& Astrophysics}
\newcommand{\nat} {Nature}

%
%


\title{Solar Cycle Precursors and the Outlook for Cycle 25}

%
%




\authors{L. A. Upton\affil{1} and D. H. Hathaway\affil{2}}

 \affiliation{1}{Southwest Research Institute, 1050 Walnut, St Suite 300, Boulder, CO 80302}
 \affiliation{2}{Hansen Experimental Physics Lab., Stanford University, HEPL-4085, Stanford, CA 94305-408, USA}





\correspondingauthor{L.A. Upton}{lisa.upton@swri.org}




\begin{keypoints}
\item Solar Cycle 25 has revealed itself as a small cycle with an expected sunspot number maximum of about 134.
\item Three magnetic precursors from the cycle minimum in 2019 predicted a similar Cycle 25 sunspot number maximum. 
\item Two of these magnetic precursors were accurately predicted years before cycle minimum using surface flux transport.
\end{keypoints}

%
%

%
%


\begin{abstract}
Sunspot Cycle 25 over 3 years past the cycle minimum of December 2019. At this point, curve-fitting becomes reliable and consistently indicates a maximum sunspot number of $135 \pm 10$ - slightly larger than Cycle 24's maximum of 116.4, but well below the Cycles 1-24 average of 179 (ranging from 81 for Cycle 6 to 285 for Cycle 19).
A geomagnetic precursor, the minimum in the $aa$-index, and the Sun's magnetic precursors, the polar field strength and axial dipole moment at the time of minimum, are often used to predict the amplitude of the cycle at (or before) the onset of the cycle.
We examine Cycle 25 predictions produced by these precursors. 
The geomagnetic precursor indicated a Cycle 25 slightly stronger than Cycle 24, with a maximum of $132 \pm 8$.
The Sun's magnetic precursors indicated that Cycle 25 would be similar to Cycle 24, with a maximum sunspot number of $120 \pm 10$ or $114 \pm 15$.
Combining the curve-fitting results with the precursor predictions, we conclude that Cycle 25 will have a maximum smoothed sunspot number of $134 \pm 8$  with maximum occurring late in the fall of 2024.
Models for predicting the Sun's magnetic field ahead of minimum, were generally successful at predicting the polar precursors years in advance.
The fact that Sun's magnetic precursors at cycle minimum were successfully predicted years before minimum and that the precursors are consistent with the size of Cycle 25 suggests that we can now \add{more} reliably predict the solar cycle.
\end{abstract}

\section*{Plain Language Summary}
Now that over 3 years have passed since the start of Cycle 25, we can determine the size of the cycle and look back at previous predictions.
For the last eleven months we have consistently found that Cycle 25 is following the behavior of a smaller than average sunspot cycle, just slightly larger than the last cycle.
The strength of the Sun's magnetic field at the start of a sunspot cycle has become recognized as the best predictor for the ultimate strength of that cycle.
This follows from solar magnetic dynamo models in which the magnetic field at minimum gets stretched and strengthened to produce the magnetic sunspots and explosive magnetic activity of cycle maximum.
Three different measurements of the strength of the Sun's magnetic field in late 2019 and early 2020 (at the start of the current sunspot cycle) indicated that this cycle would be slightly stronger than the previous cycle, but still weaker than average.
Models can be used to estimate two of these measurements well before cycle minimum, thus providing a reliable prediction years before the start of a sunspot cycle.

%
%

\section{Introduction}

Activity on the Sun varies with a periodicity of about eleven years.
This variability is characterized by fluctuations in the appearance of sunspots but also includes the evolution of coronal holes, changes in the solar wind speed, and changes in the frequency of eruptive events such as solar flares and coronal mass ejections \cite{2015Hathaway}.
Together, these and other solar phenomena drive changes in the interplanetary environment, i.e. space weather.
As space weather events interact with the Earth, they cause geomagnetic storms that impact the geospace environment in a variety of ways.
Extreme ultraviolet and x-ray emissions give rise to ionization in the ionosphere and heating and increased density in the thermosphere.
This increases the drag on satellites and debris in low Earth orbits and increases the risk of satellites colliding with debris or even completely deorbiting - as occurred in early 2022 with the SpaceX Starlink satellites \cite{2022Hapgood_etal}.
Increases in energetic charged particles can further disrupt, damage, or cause failure of satellites or their electrical components.
While satellites essential for communications and national defense are most at risk, geomagnetic storms are not a threat to just them.
Radiation from these storms also pose a threat to astronauts in space and crew on airline flights over the poles.
Closer to home, ground-induced currents produced by geomagnetic storms can overwhelm power grids, resulting in power outages.
Predicting the size of a solar cycle is an important step in improving our ability to prepare for space weather years in advance. 

Two (related) physics based precursors for predicting the amplitude of a solar cycle have risen to the top as being the most reliable: geomagnetic activity levels and the Sun's magnetic configuration (polar fields and axial dipole moment) near the time of sunspot cycle minimum.
Cycle predictions can be made well before cycle minimum by using models to simulate the evolution of the Sun's surface magnetic field.
Once the cycle is well under way, curve fitting can be used to determine the amplitude of the cycle.
In this paper, we present a comprehensive picture of the outlook for Solar Cycle 25 based on each of these methods, individually and as a combined prediction.

\section{Curve Fitting and the Amplitude (Maximum) of Cycle 25}

Cycle 24/25 minimum occurred in December of 2019 and we are now over three years into Cycle 25. At this stage of the solar cycle, curve fitting becomes quite reliable. 
\citeA{1994Hathaway_etal} proposed a parametric curve for fitting to the monthly sunspot numbers, $R$, in each cycle with:

\begin{equation}
    R(t) = A (t - t_0)^3 [\exp[(t -t_0)^2/B^2] - C]^{-1}
\end{equation}

\noindent where $R$ is the relative sunspot cycle number and $t- t_0$ is the time in months since the effective start of the cycle.
Starting with four-parameters (amplitude $A$, rise time $B$, asymmetry $C$, and starting time $t_0$), they showed that the asymmetry parameter $C$ could be fixed \add{as 0.71} for all cycles and the rise time parameter $B$ could be expressed in terms of amplitude \remove{(this is the Waldmeier effect -- big cycles rise more rapidly to maximum)} 
\add{with:}
\begin{equation}
    B = 27.12 + 25.15.72/(A \times 10^{3})^{1/4} 
\end{equation}
\add{(This is the Waldmeier effect -- big cycles rise more rapidly to maximum. By directly using the rise rate,} \citeA{2022Pawan_etal} \add{predicted a $R_{max}(25) = 138 \pm 26$.)} 
\add{These refinements to the fitting function allow} the cycle to be fit with just two parameters - the starting time for the cycle, $t_0$, and the amplitude, $A$. 

In 2015, the sunspot number was revised to version 2.0 \cite[hereafter, V2.0]{2016Clette_etal_SoPh, SSN}\add{, which increases the sunspot numbers by changing the multiplicative factor from 0.6 to 1.0 and by using the average of multiple observers}.
We find that the values for $B$ and $C$ have changed for V2.0, with the best fit for B and C  now given by:

\begin{eqnarray}
  B &=& 36.3 + 0.72/\sqrt{A} \ \rm{months} \\
  C &=& 0.70 
\end{eqnarray}

\begin{figure}
 \centerline{\includegraphics[width=0.76\textwidth,trim={10 08 10 10},clip]{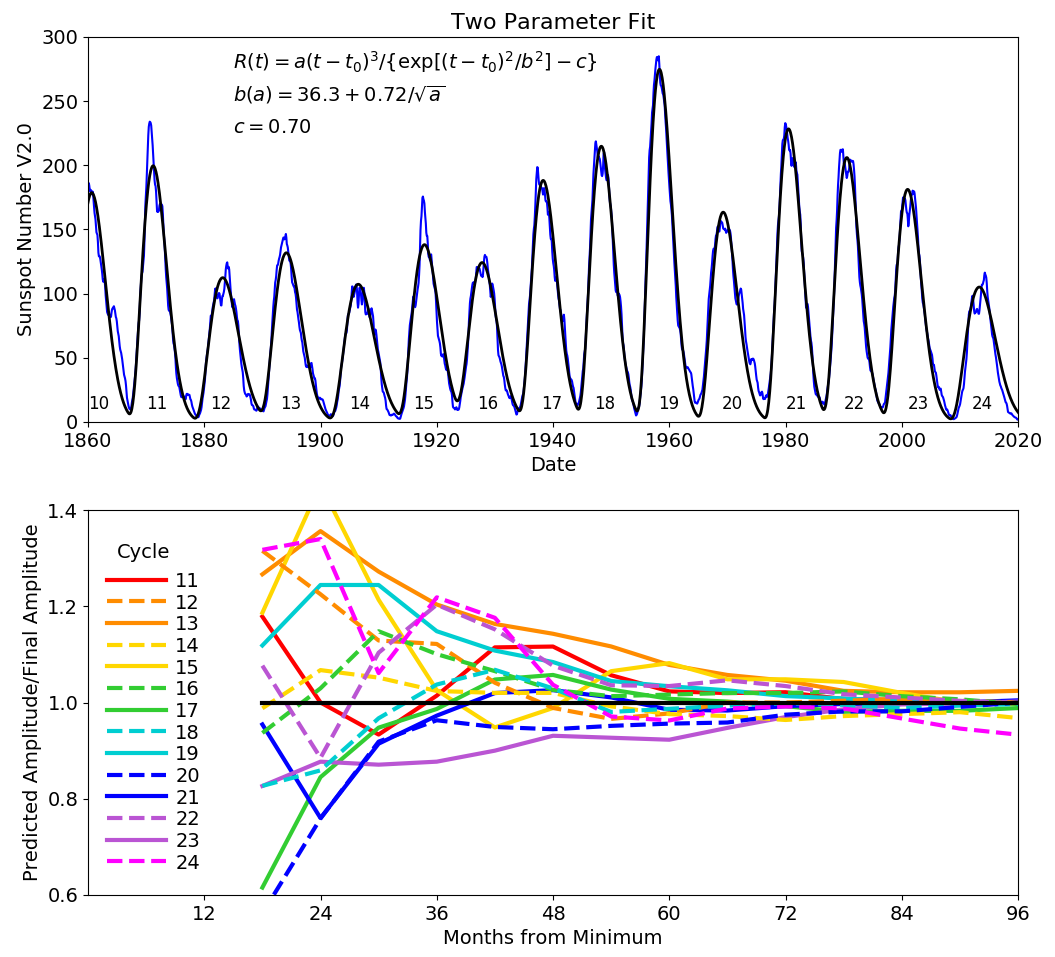}}
\caption{The two parameter functional fit to each of the last 14 cycles (top).
The fit amplitudes relative to their final values as functions of time lapsed since minimum for the last 14 cycles (bottom).
}
\label{fig:FitPerformance}
\end{figure}

\noindent We apply a Levenberg-Marquart method \cite{1992Press_etal} to fit this nonlinear function of two parameters to the monthly sunspot numbers.
Using the uncertainties in the monthly sunspot numbers, we generate 100 Monte Carlo realizations of the monthly sunspot number record, which are then fit to determine the uncertainties in the fit parameters.
The goodness of the fit between this two parameter function and the smoothed sunspot number for previous cycles is shown in the top panel of Figure~\ref{fig:FitPerformance}. 

We investigate the accuracy of the fitting as the cycles progress by performing the curve fitting at six month intervals from each cycle's minimum. We then compare the amplitude of the fit at each interval to the observed final amplitude fit.
We find that using the the two parameters, the interval fit typically converges to within 10\% of the final fit values after about 3 years into each cycle (near the inflection point on the rising curve), as illustrated in the bottom panel of  Figure~\ref{fig:FitPerformance}.
Note that there is a tendency to overestimate the amplitude early in each cycle. 

The curve fitting results for Cycle 25 were variable and quite uncertain until a year ago (May 2022 -- 30 months after minimum).
Since then the results have been virtually unchanged - the curve fitting has given $R_{max}(25) = 135 \pm 10$ with $t_0 = 2019.8 \pm 0.3$.
This determination of the amplitude of Cycle 25 is unlikely to change by more than 10\% as the cycle continues to develop.

\section{Magnetic Precursors for Predicting Cycle Amplitudes}

\begin{figure}
 \centerline{\includegraphics[width=0.8\textwidth]{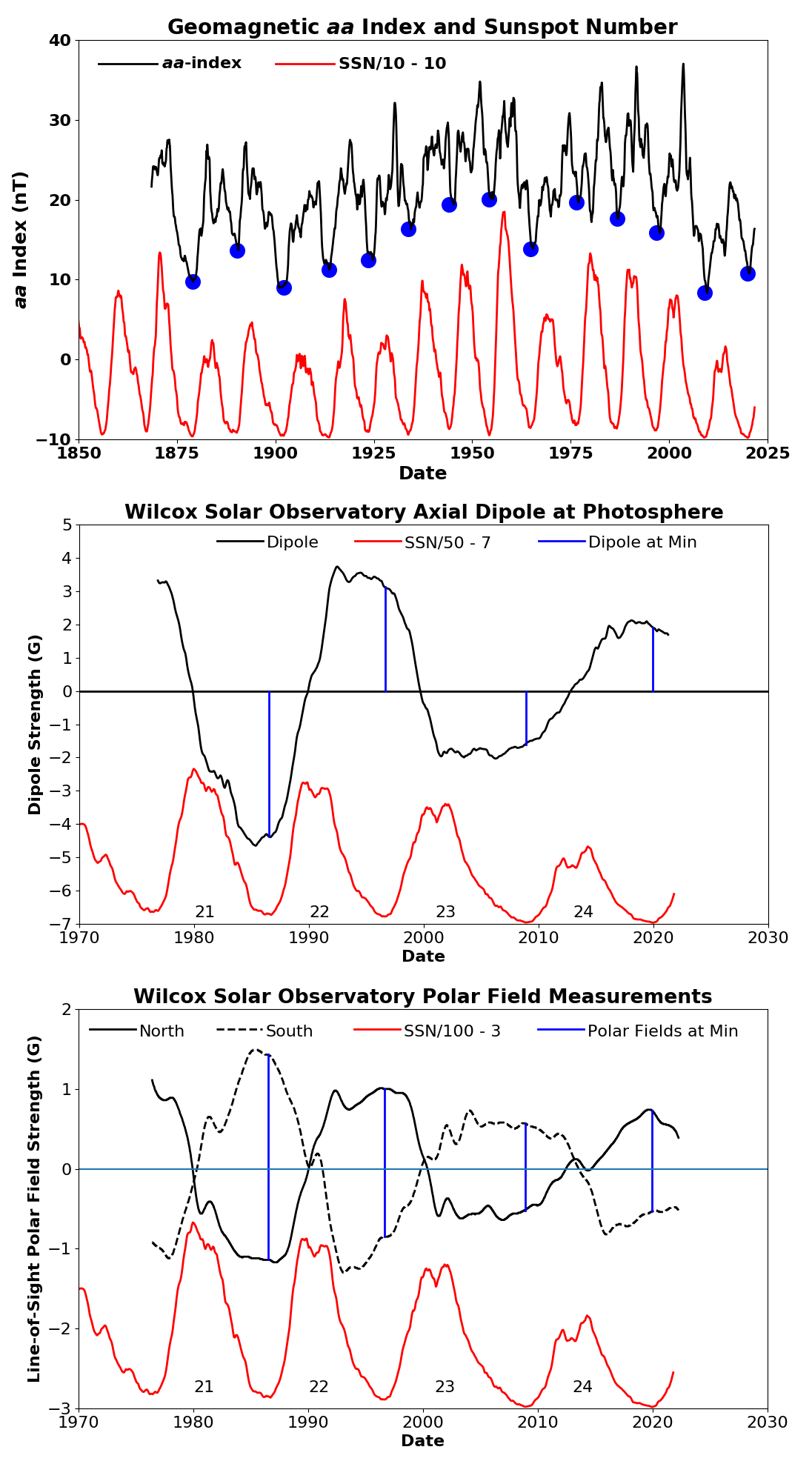}}
\caption{Solar cycle prediction precursors from magnetic conditions (in blue) near cycle minimum.
Top panel: The geomagnetic $aa$-index in black with minimum values in blue.
Middle panel: The Sun's axial dipole strength from WSO in black with values at cycle minima in blue.
Bottom panel: The Sun's polar fields from WSO in black with values at cycle minima in blue.
The smoothed sunspot number (V2.0) is shown in red in each panel.
}
\label{fig:SSN}
\end{figure}

Geomagnetic precursors to the solar cycle were first suggested by the Russian researcher Ohl in 1966 \cite{1979ohl}.
He found that the minimum in geomagnetic activity, as indicated by the \textit{aa} index \cite[plotted in the top panel of Figure~\ref{fig:SSN}]{aaIndex}, is strongly correlated with the amplitude of the oncoming cycle .
This can be seen by comparing the heights of the minima, shown in blue, to the following peaks in sunspot number, shown in red.
(Note that we include a correctional offset of +3 nT prior to 1957, when the English observing station was moved from Abinger to Hartland \cite{2005SvalgaardCliver}.)
A slight drawback to this method comes from the fact that these geomagnetic minima usually occur just after the sunspot cycle minimum.
Other geomagnetic precursor methods have been devised to provide earlier predictions but are either based on data that don't cover as many sunspot cycles  or require data processing with free parameters \cite{1982Feynman, 1993Thompson}.

The Sun's polar magnetic field configuration near the time of cycle minimum is gaining popularity as a precursor for the amplitude of the following cycle.
At minimum, the Sun's magnetic field is well characterized by a simple axial dipole.
This simple configuration is often the starting point for models of the Sun's magnetic dynamo \cite{1961Babcock}.
The geomagnetic precursors near minimum are thought to perform well because they are driven by high-speed solar wind streams and are thus reflections of the strength of the Sun's polar magnetic fields \cite{1987SchattenSabatino, 2009WangSheeley}.
Measurements of the Sun's polar fields over the last four solar cycles have proven to be successful predictors of the following cycle amplitude \cite{1978Schatten_atal, 2005Svalgaard_etal, 2010Petrovay, 2013MunozJaramillo_etal}.

Direct, systematic (daily) measurements of the Sun's polar fields have been made at the Wilcox Solar Observatory (WSO)  since 1976 \cite{WSO, 1995Hoeksema,1978Svalgaard_etal}.
The polar magnetic field configuration can be characterized in two different ways: by calculating the axial dipole component of the Sun's magnetic field (Figure~\ref{fig:SSN}, middle panel) or by averaging the flux density over each polar region (i.e., the polar field strength, shown in Figure~\ref{fig:SSN}, bottom panel).
The latitudes and field component (e.g., radial or line-of-sight) used to calculate the flux density at each pole are somewhat arbitrary.
For WSO, the polar field measurement was set by the spatial resolution of the instrument and defined using the highest latitude pixel, which measures the line-of-sight fields nominally between 55$^\circ$ and the poles.
Magnetic data from other instruments have often employed the radial component and used different latitude ranges. While the flux density over each polar region offers insight into hemispheric asymmetries, the innate ambiguity associated with this measurement may make the axial component of the Sun's magnetic dipole a better metric for solar cycle prediction \cite{upton2014a}.

\section{Magnetic Precursor Measurements at Cycle 24/25 Minimum}

We now have observations of the Sun's polar fields as well as measurements of geomagnetic activity during (and for three years after) the Cycle 24/25 minimum.
For the geomagnetic precursor, we focus on the minimum in geomagnetic activity as measured by the \textit{aa} index (Ohl's method) because it does not require any decomposition and the measurements date back to 1868, allowing the relationship to be determined for 13 cycles.
For the polar field precursor, we look at both the polar field strength and the axial dipole strength during solar cycle minimum, as measured by WSO.
These measurements are all shown in Figure~\ref{fig:SSN}, along with the smoothed sunspot number V2.0.
While the polar field measurements do indeed appear to be indicative of the strength of the next cycle, they only provide three solar cycles (Cycles 22-24) for determining the relationship between the polar fields at minimum and the amplitude of the next cycle. 

 \begin{figure}
 \noindent\includegraphics[width=\textwidth]{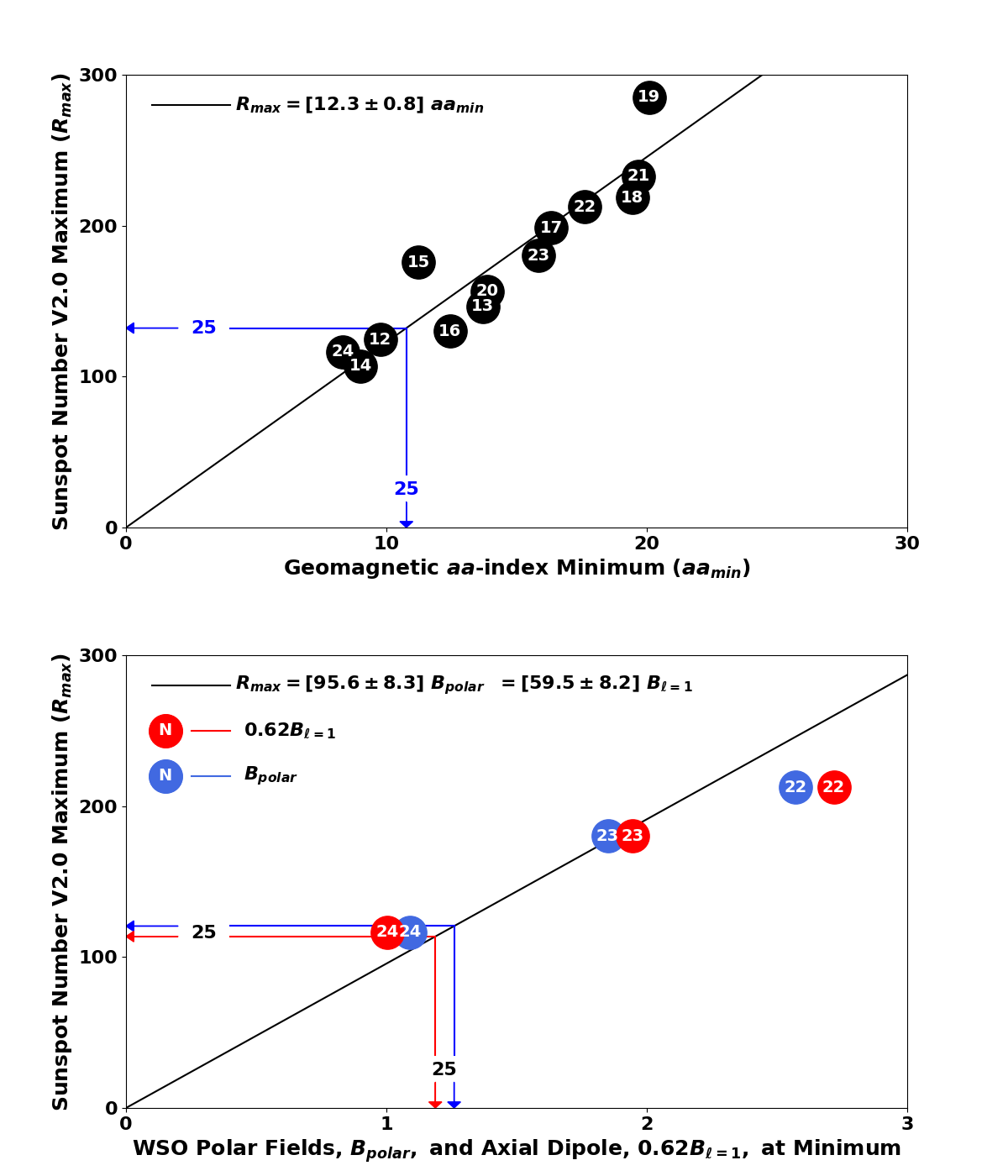}
\caption{Sunspot number maxima as functions of the magnetic precursors.
Top panel: Cycle maxima plotted as functions of the minima in the geomagnetic $aa$ index.
Bottom panel: Cycle maxima as functions of the WSO axial dipole at the photosphere in red and the WSO polar fields in blue.
In both panels the average ratio of sunspot number maximum to precursor value is given by the black line.
The values of the precursors at Cycle 24/25 minimum are shown by the blue and red vertical lines while the values they indicate for Cycle 25 maximum are shown by the horizontal lines.
}
\label{fig:precursors}
\end{figure}

We begin by relating each precursor measurement to the strength of following cycle as indicated by $R_{max}$, i.e., the maximum smoothed sunspot number V2.0 for that cycle.
This is shown in Figure~\ref{fig:precursors}. 
We calculate the ratio of $R_{max}$ to the minimum of the geomagnetic \textit{aa} index for each cycle and find:

\begin{equation}
  R_{max}  =  (12.3 \pm 0.8) aa_{min} 
\end{equation}

\noindent where, $aa_{min}$ is the minimum in the \textit{aa} index near the time of solar minimum.
Likewise, We calculate the ratio of $R_{max}$ to the polar fields and axial dipole moment and find: 

\begin{eqnarray}
  R_{max} &=&  (95.6 \pm 8.3) B_{polar} \\
  R_{max} &=&  (59.5 \pm 8.2) B_{l=1}
\end{eqnarray}

\noindent where, $B_{polar}$ is the absolute value of the difference between the north and south polar field strengths  and $B_{l=1}$ is the axial dipole field strength. 
While all three methods do suggest a weak cycle, they don't completely agree.
The minimum in aa-Index of 10.76  gives $R_{max}(25) = 132 \pm 8$, the WSO polar fields of 1.26 give $R_{max}(25) = 120 \pm 10$, and the WSO axial dipole of 1.91 gives $R_{max}(25) = 114 \pm 15$.
The variance weighted mean of these precursors gives $R_{max}(25) = 125 \pm 6$ where reported errors are 1$\sigma$ errors, which is slightly lower that, but within the range given by the current curve fitting values.

 \section{Combined Prediction}

Both the magnetic precursors and the curve fitting indicate that Cycle 25 sunspot number maximum will be slightly bigger than the 116 of Cycle 24 but significantly smaller than the average of 179 for all 24 previous cycles\add{, which ranged from 81 for Cycle 6 to 285 for Cycle 19}.

\begin{figure}
 \noindent\includegraphics[width=\textwidth]{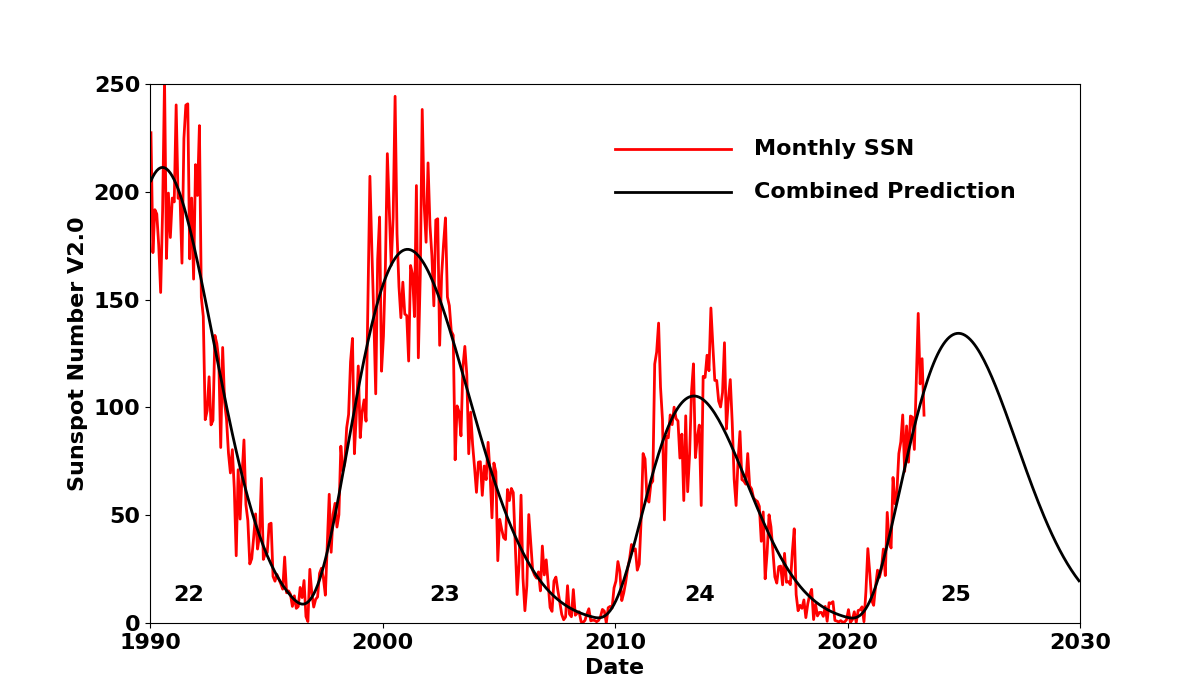}
\caption{Combined sunspot number prediction based on the magnetic precursors and curve fitting at 40 months into Cycle 25.
Monthly sunspot numbers V2.0 are noisy and shown in red.
The prediction curves are smooth and shown in black.
}
\label{fig:SC25_new}
\end{figure}

\cite{1999Hathaway_etal} proposed a combined prediction based on a weighted mean of the precursor predictions and the curve fitting prediction with more weight given to the curve fitting as the cycle progresses (50/50 at 36 months past minimum). 
This combined prediction gives a maximum smoothed sunspot number of $R_{max}(25) = 134 \pm 8$  with maximum occurring late in the fall of 2024.  
Figure~\ref{fig:SC25_new} shows a plot of the predicted curve, along with the observed monthly sunspot numbers V2.0.

\section{Magnetic Precursor Estimates before Minimum}

Modelers have sought to extend the predictive range of the polar field precursors by using physics based dynamo models \cite{2020Charbonneau, 2021Nandy} or Surface Flux Transport (SFT) models \cite{2005Sheeley, 2014Jiang_etalB} to obtain the polar fields years ahead of minimum. \citeA{2022Pawan_etal} \add{showed that the rise rate of the polar fields during the first three years after the reversal gave similar results as the rise rate of the cycle, potentially providing an avenue for modelers to further extend their predictive range.}
Before cycle minimum in 2019, the Solar Cycle Prediction Panel, which represents NOAA, NASA and the International Space Environmental Services (ISES), released its official sunspot number forecast \cite{NOAA}, predicting a sunspot number maximum of $115 \pm 10$. Their consensus forecast was based largely on the model projections for the polar fields at minimum.

In 2016 and 2018, we used our Advective Flux Transport (AFT) model to create forecasts of the Sun's polar fields in order to predict the strength of Solar Cycle 25 \cite{2016HathawayUpton, 2018UptonHathaway}.
Our results indicated that the polar fields at Cycle 24/25 minimum would be similar to those at the previous minimum, suggesting Solar Cycle 25 would be a weak cycle, very similar in amplitude to Solar Cycle 24.
Another SFT model  \cite{2016Cameron_etal, 2018Jiang_etalJ} reported similar results, predicting that Solar Cycle 25 would be slightly larger than Cycle 24 \add{(with a maximum of 116.4)}, but less than the more average sized Cycle 23 \add{(with a maximum of 180.3, very close to the Cycle 1-24 average of 179)}. 
  
\begin{figure}
 \noindent\includegraphics[width=\textwidth]{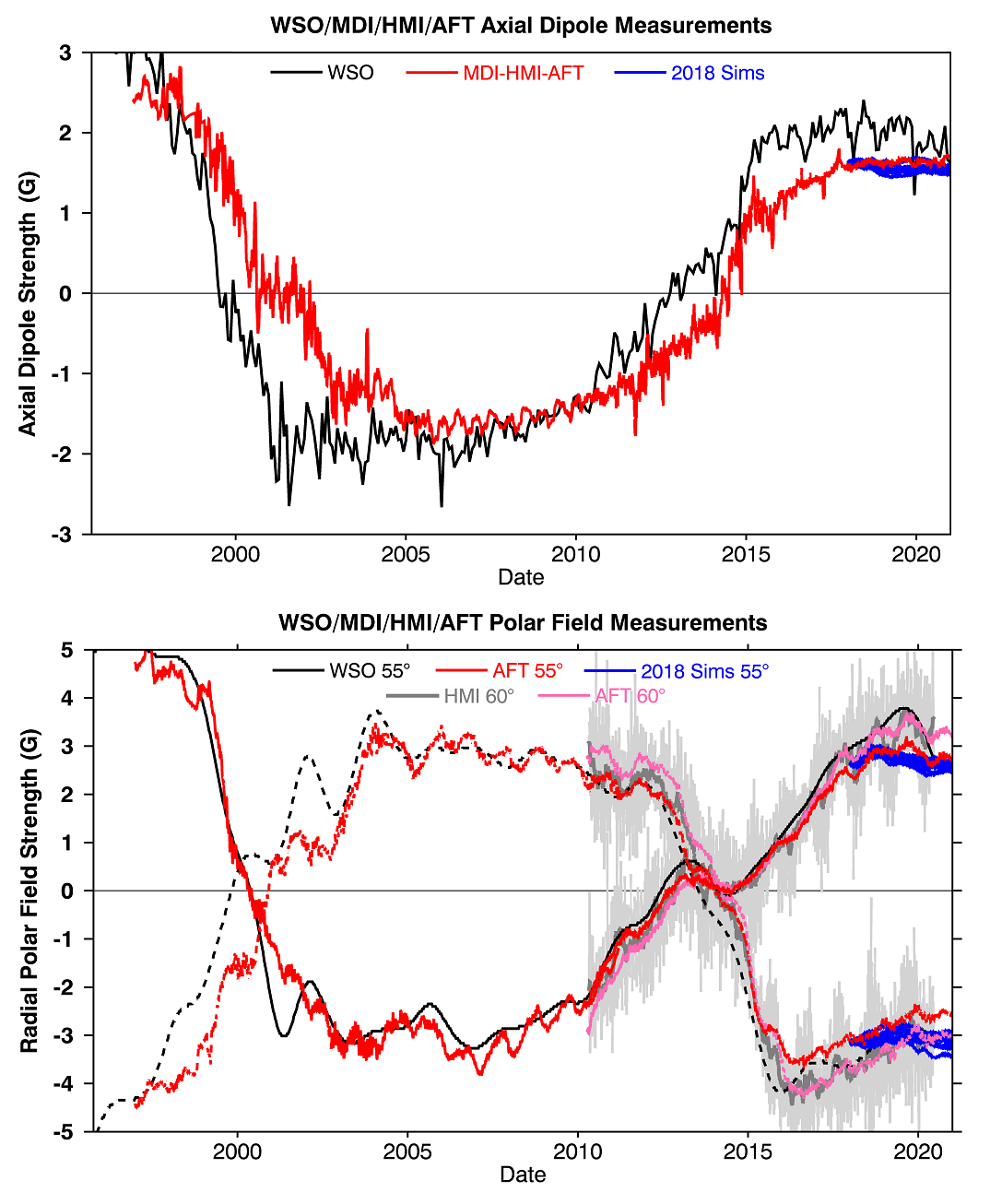}
\caption{Polar precursor measurements. This figure shows the axial dipole measurements  in the top panel and the hemispheric polar fields strengths in the bottom panel. WSO \add{$B_{l=1}$} is shown in black, MDI/HMI/AFT measurements are shown in red/pink (55$^\circ$/60$^\circ$), and the \cite{2018UptonHathaway} predictions are shown in blue. The northern/southern polar fields are are shown with a solid/dashed line. For additional reference, the HMI radial polar field measurements are shown in light gray (smoothed in dark gray). NOTE: WSO measurements have been re-scaled to be more consistent with the HMI strengths\add{: the axial dipole $B_{l=1}$ was multiplied by a spherical harmonic normalization factor of 2.064 and the polar fields were multiplied by a LOS to radial field conversion factor of 5.11}. }
\label{fig:AFT}
\end{figure}

The AFT model assimilates the observed magnetic field from the SOHO/MDI \cite{1995Scherrer_etal} and SDO/HMI \cite{2012Scherrer_etal} magnetograms in order to provide the closest contact with the observations.
This mode, known as the AFT Baseline, provides the most accurate representation of the magnetic field over the entire surface of the Sun \add{based solely on near-side observation}.
In it's predictive mode, AFT starts with a map from the AFT Baseline and advances it forward to make predictions about how the Sun's magnetic field will evolve.
In 2016 and 2018, we used AFT to create predictions of the Sun's polar field evolution in order to predict the strength of Solar Cycle 25. Now that we are well past minimum, and Cycle 25 is well under way, we revisit those predictions to assess their performance. 

In \cite{2016HathawayUpton}, we predicted an axial dipole for the solar minimum of $1.36 \pm 0.20$ G and in \cite{2018UptonHathaway} we predicted an axial dipole of $1.56 \pm 0.05$ G. (A comparison of those results and detailed discussion can be found in \cite{2018UptonHathaway}.)
We show our 2018 polar field predictions in Figure~\ref{fig:AFT} (blue lines), along with the corresponding measurements from WSO (in black), and the AFT Baseline constructed with the MDI, and HMI observations (in red).
AFT was able to successfully predict the evolution of the AFT/HMI axial dipole. However, the axial dipole in the WSO observations was about 20\% higher than HMI and AFT at the time of cycle minimum.

The AFT  polar field strengths (bottom panel of Figure~\ref{fig:AFT}) are calculated from the radial magnetic field above 55$^\circ$ (red line) and 60$^\circ$ (pink line). The LOS WSO measurements (nominally 55$^\circ$ and above) are shown in black and the HMI derived Mean Radial Fields 60$^\circ$ \cite{HMIpolar, 2015Sun_etal} are shown in light gray (smoothed values in dark gray). 
AFT was able to successfully predict the evolution of the AFT/HMI polar fields in the North, but that the polar fields in the South did diverge somewhat starting in late 2019.
The difference between the north and south polar field strengths in the 2018 predictions ranged from 5.49 to 5.92, with an average of 5.68. The observed difference was 5.25, or about 10\% smaller than the AFT 2018 ensemble of predictions. Note that the radial AFT/HMI polar fields above 55$^\circ$ are smaller than the WSO single pixel LOS polar fields corresponding to the same latitude range.
The AFT Baseline polar fields above 60$^\circ$ at the start of 2020 agree with the WSO fields above 55$^\circ$ and with the HMI derived fields above 60$^\circ$.

\section{Discussion}

Geomagnetic and solar magnetic precursors have been shown to be most reliable predictors of an impending solar cycle.
However, there are still some uncertainties associated with these predictors.
Currently, the geomagnetic precursors are more robust due to the length of the dataset, but the physical mechanism behind their success is less direct.
The polar precursors have a stronger foundation in physics, but with a much shorter time line the functional relationship is poorly defined.
This is confounded by the fact that the polar measurements, in and of themselves, are not well constrained. This is primarily due to innate observational limitations.
This is highlighted in Figure~\ref{fig:AFT} by a) the mismatch between the WSO and the MDI/HMI/AFT axial dipole measurements  (top panel), b) the annual oscillation in the WSO LOS polar field measurements (bottom panel), and  c) the spread in the HMI radial polar field measurements (bottom panel).

The WSO axial dipole and the MDI/HMI/AFT axial dipole appear to be offset in both time and amplitude.
The offset in time is most apparent during the dipole reversals, as the WSO reversals precede the MDI/HMI/AFT reversal by about a year or two. The offset in amplitude is most apparent during last solar minimum, when the axial dipole is relatively flat. These offsets are a consequence of the limited resolution of the WSO observations convolved with the changing latitude range of the WSO pixel (due to an orbit around the inclined Sun). As previously mentioned, the highest latitude WSO pixel measures the line-of-sight fields nominally between $55^\circ$ and the poles. However, the Sun's rotation axis is inclined $\sim 7.15^{\circ}$ with respect to the ecliptic plane. So, while on average the highest latitude pixel measures from $55^\circ$ and above, the latitudes actually being measured actually vary between $48^\circ$ and above to $62^\circ$ and above. This produces a seasonal oscillation in the measurements of the Sun's polar fields, which is clearly visible as an annual signal in the plot of the hemispheric polar field strengths (bottom panel of Figure~\ref{fig:AFT}). Furthermore, the northern/southern (solid/dashed) hemisphere measurements are 6-months out of phase with one another, such that the northern/southern WSO measurements come into better agreement with the MDI/HMI/AFT measurements in the Spring/Fall, when the the Southern/Northern pole is inclined toward the Earth. It should be noted that these annual signals are present in the MDI and HMI data as well, though to a lesser extent. This is because the higher resolution of MDI and and HMI provide more detailed latitudinal coverage, but the inclined orbit causes the line-of-sight angle of the magnetic field (with respect to the radial component) to also change by  $\pm 7^\circ$ over the course of the yearly orbit.

The deviations in the polar field measurements are most notable when the polar fields are rapidly evolving, during the polar field reversals.
At this time, new polarity flux is being transported to the poles to cancel with the old polarity polar flux, causing a large latitudinal gradient in the high latitude flux. As the Earth orbits the Sun and the latitudes measured by the most poleward WSO pixel change, the pixel samples lower/higher latitudes and more of the new/old polarity flux is present in that pixel. Consequently, the average polar field strength measured by that pixel changes substantially over the course of the orbit resulting in the observed  annual oscillation. 

These deviations in the hemispheric polar field strengths feed into the measurements of the axial dipole moment (top panel of Figure~\ref{fig:AFT}). However, rather than appearing as an annual oscillation, they instead present as the offsets in amplitude and time noted above. One might expect that the deviations would cancel out, since the northern and southern hemispheres are out of phase with one another and one of the poles is always favorable. However, the unfavorable pole always appears ``ahead'' in the polar reversal process (because the lower latitudes that are being include have more new polarity flux). This means that at nearly all times, one of the two poles will seem to be ``ahead'' and thus the evolution of the axial dipole as measured by WSO will precede the true polar reversal in time, producing the apparent temporal offset seen in the top panel of Figure~\ref{fig:AFT}. This is evidenced by the fact that the WSO polar field measurements preferentially precede the MDI/HMI polar field measurements.

If not properly accounted for, the temporal deviations can also produce an offset in amplitude. 
The WSO axial dipole moment during minima is the ``yardstick'' used to determine the strength of the next cycle for prediction purposes.
Since, the axial dipole moment is most crucial during solar minima, one might naturally scale the the WSO and MDI/HMI measurements such that they are in agreement during the Solar Cycle 22/23 and Solar Cycle 23/24 minima (1999 and 2009), as done in Figure~\ref{fig:AFT}. However, doing so results in the WSO dipole being about 20 percent higher than the HMI dipole for the Solar Cycle 24/25 minimum in December 2019. This faulty scaling resulted in the AFT solar cycle prediction underestimating the strength of Solar Cycle 25, not because the flux transport did not accurately predict the evolution of polar fields, but rather because the ``yardstick'' used to make the prediction was not calibrated properly.      

While outside the scope of this paper, a detailed accounting of these deviations and a proper calibration of the WSO observations with respect to the modern space observations is needed in order to reduce the uncertainty in the polar field-SSN relationship.
Uncertainty in the polar measurements and the polar field-SSN relationship would also greatly benefit from polar mission to measure the magnetic fields from directly overheard thus removing uncertainty associated with LOS observations of the magnetic field.

Uncertainty in the polar measurements not withstanding, the predictions presented in this paper firmly point to Solar Cycle 25 being slight larger than Solar Cycle 24, but no where close to the amplitude of Solar Cycle 23 or the average solar cycle strength ($\sim 180$). This will be the first increase to the cycle amplitude that we've seen since solar Cycle 21 (e.g., about 50 years). This may mean that we have reached the inflection point in the current Gleissberg cycle and might start to see bigger cycles again.

\section{Open Research}


The results presented in this paper rely on geomagnetic indices \cite{aaIndex} calculated and made available by ISGI Collaborating Institutes from data collected at magnetic observatories.
We thank the involved national institutes, the INTERMAGNET network and ISGI (\url{isgi.unistra.fr}). 
Wilcox Solar Observatory data \cite{WSO} used in this study was obtained via the web site \url{http://wso.stanford.edu/Polar.html} courtesy of J.T. Hoeksema. The Wilcox Solar Observatory is currently supported by NASA. HMI Polar Field data 
\cite{HMIpolar} used in this study was obtained via the web site \url{http://jsoc.stanford.edu/ajax/lookdata.html?ds=hmi.meanpf_720s}. HMI is currently supported by NASA.
SILSO sunspot number data \cite{SSN} was obtained via the web site \url{https://www.sidc.be/silso/} courtesy of the Royal Observatory of Belgium, Brussels.


\acknowledgments

LAU was supported by NASA Heliophysics Living With a Star grants NNH16ZDA010N-LWS and NNH18ZDA001N-LWS and by NASA grant NNH18ZDA001N-DRIVE to the COFFIES DRIVE Center managed by Stanford University.
DHH was supported by NASA contract NAS5-02139 (HMI) to Stanford University.
HMI data used in this study are courtesy of NASA-SDO and the HMI science team. MDI data used in this study are courtesy of NASA/ESA-SOHO and the MDI science team.
We thank Todd Hoeksema, Marc DeRosa, Xudong Sun, and Andres Munoz-Jaramillo for useful discussions regarding the uncertainty in the polar field measurements. \add{We also thank the anonymous referees whose useful suggestions improved the overall quality of the paper.}
 

%
%




\bibliography{main}

\end{document}